# BALANCING PHYSICAL MODELING AND MUSICAL REQUIREMENTS: ALGORITHMICALLY SIMULATING THE CALLS OF HYALESSA MACULATICOLLIS FOR REAL-TIME INSTRUMENTAL CONTROL

**Staas de Jong** (`staas@freedom.nl`) [1]

[1] **Universiteit Leiden**, *Honours Academy*, Leiden, NL

## ABSTRACT

This paper presents an algorithm that simulates the calls of the Hyalessa maculaticollis cicada for musical use. Written in SuperCollider, its input parameters enable real-time control of the insect call phase, loudness, and perceived musical pitch. To this end, the anatomical mechanics of the tymbal muscles, tymbal apodeme, tymbal ribs, tymbal plate, abdominal air sac, tympana, and opercula are physically modeled. This also includes decoherence, following the hypothesis that it, in H. maculaticollis, might explain the change in timbre apparent during the final phase of a call sequence. Overall, the algorithm seems to illustrate three main points regarding the trade-offs encountered when modeling bioacoustics for tonal use: that it may be necessary to prioritize musical requirements over realistic physical modeling at many stages of design and implementation; that the resulting adjustments may revolve around having physical modeling perceptually yield sonic events that are well-pitched, single-attack, single-source, and timbrally expressive; that the pitch-adjusted simulation of resonating bodies may fail musically precisely when it succeeds physically, by inducing the perception of different sound sources for different pitches. Audio examples are included, and the source code is structured and documented so as to support the further development of cicada bioacoustics for musical use.

## 1. INTRODUCTION

Reflecting on immersive sound, music and computing – the theme of this year's conference – one way in which forms of instrumental control can be designed so as to recognizably tie musical pieces in which they are used to specific geographical regions, time periods, or circumstances, is to have the heard sounds that they produce mimic those that occur in the corresponding, natural environments.

Simultaneously, however, another musical requirement may well be that these heard sounds also should have a clearly perceived pitch which can be changed in real time – e.g. to enable melodies; and that across such pitch or other control changes, the resulting heard sound will still, via the process of human Auditory Scene Analysis (ASA, see Bregman in [1]), be perceived as coming from a single source – e.g. to help enable use in counterpoint.

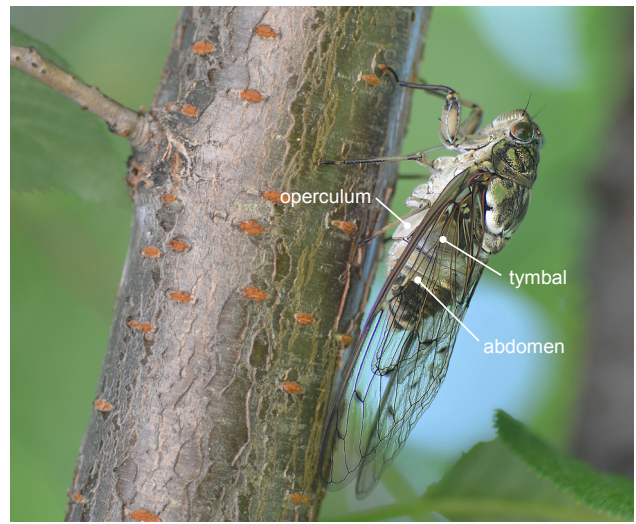

Figure 1. An H. maculaticollis individual in its natural environment, showing its tymbal, abdomen & operculum. [2]

This paper presents an algorithm that attempts to balance these three musical requirements for the case of H. maculaticollis, a species of cicada that is native to the Korean Peninsula.

Admittedly, when looking for a suitable starting point to investigate physical modeling versus musical requirements, choosing tonal use may not immediately seem like an obvious choice. After all, are there not more recently introduced musical paradigms available, subject to more dynamic ongoing developments?

However, if we do not want to first have to debate the contents of a specific musical paradigm, and instead want to focus on its interplay with physical modeling, this seeming disadvantage becomes an advantage: The requirements for tonal use are relatively well-understood, static, and non-controversial. For example, it seems relatively clear-cut that tonal aesthetics will usually be violated if sound synthesis does not provide musical voices with accurately and precisely controlled perceived pitches.

---

[2] Cropped and annotated version of a photograph by 더 그레이스, licensed under `https://creativecommons.org/licenses/by-nc/4.0/` and obtained via [2].

## 1.1 Approach: Physical Modeling of Anatomical Mechanics, Informed by the Analysis of Field Recordings

Earlier work on synthesizing cicada sound includes [3] by Georgaki and Queiroz. Here, a different species, Cicada orni, is studied, which produces calls that in their timbre and rhythm are quite different from those made by H. maculaticollis. Still, this work clearly illustrates how, generally speaking, cicada sound synthesis can be based on the analysis of field audio recordings. For the present work, two field recordings in particular, kindly made publicly available at [4] and [5], have been used as a reference throughout development.

However, a choice was made to also base development on biological knowledge of anatomical mechanics, motivated by the potential of H. maculaticollis to provide a detailed case study of where along the causal structure of bioacoustics trade-offs between realistic modeling and musical applicability may lie.

For gaining the required biological knowledge, Pollack provided a very useful starting point in [6], giving a clear introductory overview of the anatomical components and mechanics of cicada sound production, based on earlier work by others, while placing this within a wider context of insect bioacoustics.

H. maculaticollis can then be approached as a specific instance of cicadas and their sound production, with some of the key anatomical components illustrated in Figure 1. Here, an individual of the species is shown residing in its natural environment, with the anatomical locations of its left tymbal, abdomen, and left operculum indicated. The tymbal is an organ containing ribs that can be sequentially pulled inward by the tymbal muscles, via the internal connecting tissue of the tymbal apodeme. The resulting reverberations go into other parts of the tymbal; into the air sac that is contained within the abdomen; and into the surroundings, especially via the left and right tympana. Here, each tympanum, while itself an acoustically open membrane, is variably covered by its corresponding operculum.

In order to start, based on the above general context, on the development of a concrete algorithm, the information provided by Bennet-Clark and Young in [7] and [8] has been crucial: Detailed description and measurements regarding the relevant anatomy and its mechanics and acoustics are provided, as well as a rigorously tested model of the resulting sound generation, including abdominal resonation. Although this is based on the study of other species of cicada, especially Cyclochila australasiae, results are evaluated as a model for sound production in cicadas in general, and seemed a suitable starting point when attempting to develop the algorithmic simulation of cicada calls by H. maculaticollis.

Further on, the work by Smyth and Smith reported in [9] and [10] was helpful: As a part of reported work on a novel musical instrument, two general digital filter types are mentioned that can be used when implementing the modeling of C. australasiae tymbal and abdominal resonation, respectively.

## 2. DESIGN AND IMPLEMENTATION OF THE ALGORITHM

The algorithm was written from scratch in SuperCollider, and is included in annotated source format in the Appendices. It has six input arguments, each of which can be changed in real time: `a_gate_bit`, `loudness_nrm`, `t_mae_trig`, `t_mae_retrig`, `t_mi_trig`, and `freq_hz`.

The function of audio-rate argument `a_gate_bit` is to enable graceful termination of the algorithm. This both in terms of the amplitudes of its stereo audio output, and in terms of the release of claimed computational resources. Via the control-rate argument `loudness_nrm`, the perceived loudness of algorithm output can be set, along a scale from 0.0 to 1.0.

Unlike C. australasiae, perceptually, the calls of H. maculaticollis seem to follow a three-part structure, consisting of an initial phase, a repeated middle phase, and a final phase. The arguments `t_mae_trig`, `t_mae_retrig`, and `t_mi_trig` make each of these phases, respectively, subject to real-time control: A 1.0 impulse to any of these trigger-type arguments signals the algorithm to transition to the corresponding phase of the call. Here, parts of the transliterated onomatopoeic Korean word for cicada, "maemi", are used as a mnemonic, to indicate which phase of the insect call each argument refers to. For an example of the use of these input arguments to control a complete call sequence, see the waveform at the bottom of Figure 3.

Finally, the remaining, control-rate argument `freq_hz` is used to relay, as a frequency in hertz, the intended perceived musical pitch that algorithm output should induce in the listener. Here, we encounter the first trade-off between realistic physical modeling and musical requirements: Cicada song frequencies are emphatically present at thousands of hertz, but the modeled input range for `freq_hz` was set to the octave from E4 to E5. This so that the heard pitches would still lie within one of the main ranges of tonal music: that of the soprano, the highest human vocal range.

The subsections below will give an introduction to the algorithm itself. For reasons of clarity, this will be done by traversing various parts of the insect's sound-generating anatomy in a causal order. The information on cicada anatomy and sound production that is used will be based on [7] and [8], unless explicitly stated otherwise.

### 2.1 The Tymbal Muscles: Contraction & Relaxation

During cicada song, the tymbal muscles, connected to the left and right tymbal, go through repeated cycles of contracting inward and then relaxing. In the algorithm, this was modeled in the `tymbalMuscle_trig` signal, a unipolar waveform consisting of single-sample impulses, one for each contraction cycle. During the segments of cicada song where the tymbal ribs buckle in tight synchronicity (and thereby produce a tightly coherent pulse train), each `tymbalMuscle_trig` impulse models the moment in time when contraction causes the first tymbal rib to start to buckle.

Moving on to computing the *frequency* of muscle contractions during such segments, the available information on C. australasiae presented a clear problem: Live control of the algorithm should, as discussed above, yield perceived musical pitches lying between 329 and 660 Hz. This, however, is both above the 100-200 Hz range of natural tymbal muscle contractions, and below the resulting natural song frequency of around 4300 Hz.

To resolve this, a musically-motivated choice was made to have each contraction always cause exactly 4 buckling events (see Section 2.3 for details on their nature). The idea was that, in this way, muscle contraction modeled at a mostly natural rate could cause strong overtones at the precise pitch frequency specified, with the insect song overtones then present above that (while not being perceptually dominant as the pitch). Accordingly, the starting point for the computation of the `tymbalMuscleFreq_hz` signal was a division by 4 of `freq_hz`.

Beyond this, computing `tymbalMuscleFreq_hz` was based on analysis of the H. maculaticollis field recordings. Here, a first observation was that during the initial and middle phases of ordinary call sequences, tymbal muscle frequency seemed to be at mostly higher levels associated with tightly synchronous tymbal rib buckling, to then end at lower levels during the final phase. For the algorithm, a musical choice was then made to model these lower levels so as to be a whole number of octaves lower. The idea was that this might help harmonically: First, to clearly connect the final call phase to the initially heard pitch; and second, to avoid unintended dissonances during use with other musical voices in a piece. A subsequent choice, returning to primarily considering physical modeling, was then to go a single whole octave lower, but not more, as the latter seemed to yield results sounding too similar to a general end of cicada calling.

After this, a more involved physical modeling choice was to simulate the control of tymbal muscle frequency in further detail by implementing a system of envelopes. Here, each call phase was modeled by a separate envelope, with each envelope starting or terminating based on the live input from `t_mae_trig`, `t_mae_retrig`, and `t_mi_trig`; and with all envelope output combining into a single, continuous signal.

Within the individual envelopes, the target levels, durations, and curvatures defining each envelope segment were – except when based on the octave difference mentioned above – directly derived from measurements of features present in the field recording obtained from [4]. Where possible, this was done by comparing a set of 12 adjacent harmonics that were evident in spectrograms sampled at 11025 Hz. These were visible in the 3.3-5.3 kHz range of the field recording, and in the 0.9-2.7 kHz range of algorithm output (when testing at pitch E5). The results of these measurements have been explicitly labeled and unambiguously structured in the algorithm, using a series of 15 scalar constants that define the envelopes, and 4 signal variables that contain their output.

While doing the above, a musical choice was made to *not* mimick a general downward trend seen spanning across the harmonics of repeated middle call phases in the field recording. This to help avoid algorithm output going out of tune during tonal use.

A final, speculative addition to tymbal muscle frequency control was then to give an explicit presence to the idea of it also being subject to further, possibly unknown but smaller additive effects. This was done by raising or lowering the computed frequency by a semi-normally distributed random value, updated at least once every tymbal muscle contraction. The peak amplitude for this was limited to 10 cents, however, as going higher resulted in ripples in muscle frequency harmonics that were not visible in the field recordings. Perceptually, the result then seemed to yield at most a slight variation. In any case, this extra computation can be effectively removed from the algorithm by changing a single line of code.

Example output of the live computation of `tymbalMuscleFreq_hz` during a full call sequence at pitch E5 can be seen in the waveform placed second-lowest in Figure 3. The corresponding `tymbalMuscle_trig` signal can be heard in Audio Illustration 1 of the Appendices.

## 2.2 The Tymbal Apodeme: Pull & Release

Each muscle contraction cycle, the tymbal apodeme, being the connecting tissue between a tymbal muscle and the tymbal itself, is pulled inward with variable force, and then released. Relevant results from the literature on C. australasiae here include that for tymbal rib buckling events, their phase coherence as a pulse train is supported by the mechanics of the insect's anatomy from tymbal apodeme to abdominal air sac; and that at higher apodeme pulling speeds, it appears *"that the tymbal buckling mechanism offers good coupling between the vibrations of the ribs so that the rapidly excited tymbal inevitably produces a coherent waveform"*.

This information influenced physical modeling in the algorithm via two main points of interpretation: First, that we will assume that apodeme pulling speed is related to tymbal muscle contraction frequency. As a first approximation for modeling this, in the algorithm, the first stage of computing the `tymbalApodeme_pullingSpeed_nrm` signal has been to make it directly proportional to the output of the tymbal muscle frequency envelopes.

The second point of interpretation was then that, above a certain threshold of apodeme pulling speed, tymbal mechanics will pull buckling into (the same) tight coherence. In the algorithm, this was implemented by modeling `tymbalApodeme_pullingSpeed_nrm` as a normalized value, and then clipping and scaling it to a maximum corresponding to the threshold. The concrete value for the threshold was then chosen so that `tymbalApodeme_pullingSpeed_nrm` would mostly be at its maximum level during the initial and middle phases of H. maculaticollis call sequences – the parts where in the field recordings, tymbal muscle contraction frequency mostly seemed to be at higher levels associated with tightly synchronous tymbal rib buckling. The resulting computation of `tymbalApodeme_pullingSpeed_nrm` is illustrated by the waveform placed third-lowest in Figure 3.

### 2.3 The Tymbal Ribs: In-Buckling & Out-Buckling

*2.3.1 Modeling the Individual Tymbal Rib Buckling Events*

The literature indicates that for C. australasiae, during normal song, a single pull-and-release cycle by the tymbal muscle and apodeme results in the first 2, sometimes 3 tymbal ribs sequentially buckling inward, from posterior to anterior (while the $4^{th}$ rib only buckles during "protest song"); and that after this, there is a simultaneous buckling back outward.

For the algorithm, as a trade-off between physical modeling and musical requirements, the choice was made to always model 4 buckling events per muscle contraction (this has been motivated in Section 2.1): the in-buckling of tymbal ribs 1-3, followed by their combined out-buckling. The associated acoustic buckling pulses were then modeled using one envelope per buckling event, with each event represented by a bipolar, single-pulse waveform. The amplitude levels, durations and curvatures of the envelope segments defining these waveforms were based on the acoustic measurements presented in [7] in Figures 3, 4, and 7, and Table 2.

During normal song by C. australasiae, there are 2 or 3 whole cycles of the tymbal plate resonant wave (see Section 2.4) in between the buckling pulses (see e.g. Figure 5a in [7]). In the algorithm, the overall time between the start of consecutive buckling events already was to be fixed as the reciprocal of the currently intended musical pitch, as indicated via the `freq_hz` input signal. Within this context, the duration of each buckling pulse was then scaled so as to allow for precisely 2 in-phase echoes before the start of the next pulse.

Regarding the pulses' amplitudes, the overall loudness of tymbal rib buckling events may be subject to further anatomical control: In [11], it is noted for other cicada species that contraction of a tensor muscle *"increases the convexity of the tymbal and therefore the power required to make it click"*. This then causes the tymbal ribs to produce a louder sound once they do buckle. In the algorithm, the effect of this is mimicked by linear amplitude scaling, which was made subject to live (musical) control via the lower half of the input range of `loudness_nrm`.

Figure 2 shows an example of a complete sequence of tightly synchronized tymbal rib buckling pulses as computed for pitch E5 by the algorithm.

*2.3.2 Modeling Decoherence*

So far, we have already recapitulated a crucial part of how it is that the anatomy of a singing cicada can transform individual tymbal rib clicks into a longer, continuously oscillating pressure wave: The tymbal mechanics of muscle contraction, apodeme pulling, rib buckling, and plate reverberation cause a series of acoustic pulses that are tightly in phase with eachother. By "decoherence", then, we will here mean the loss of this phase coherence.

For the algorithm, we will follow the hypothesis that in H. maculaticollis, decoherence might explain the change in timbre that is apparent during the final part of a call sequence. In spectrograms of the field recordings (for an example, see the top of Figure 3) it is visible how, during the initial and middle call phases, the sound intensities at frequencies around insect song frequency are distributed across separate harmonics; while during the final phase, these intensities are still relatively higher than elsewhere in the spectrum, but much more evenly distributed. This can, of course, also be heard.

The literature indicates that, for C. australasiae, a loss of phase coherence is associated with slower apodeme pulling speed resulting in greater intervals between rib buckling events [7]. For the algorithm, it was assumed that a similar association would exist for H. maculaticollis. As a first approximation of concretely modeling this, the duration of intervals between rib buckling events was made inversely proportional to apodeme pulling speed.

However, the proportional relationships modeled so far – i.e. between tymbal muscle contraction frequency, apodeme pulling speed, and rib buckling intervals – would then not suffice, still yielding clearly separate harmonics during the final call phase. Therefore, as a first approximation of modeling the effects of further unknown processes, at each tymbal muscle contraction, the pre-buckling interval duration added by decoherence was drawn, according to a uniformly random distribution, from none up to the proportionally computed amount.

An example of the resulting algorithm output for pitch E5 can be compared to one of the field recordings in the rightmost part of the spectrograms at the top of Figure 3, starting vertically above the `mi_trig` impulse in the graph at the bottom. Also, the underlying, internally computed `tymbalRibs_buckling_sig` signal can be heard in Audio Illustration 2 of the Appendices.

### 2.4 The Tymbal Plate: Resonant Vibrations

As is described in the literature on C. australasiae, when a tymbal rib buckles, this also rapidly moves inward another part of the tymbal organ: the tymbal plate. This plate is also connected to the dorsal pad, another component of the tymbal made of a rubber-like protein called resilin. The overall mechanical result is that a buckling event will set the tymbal plate into a series of resonant vibrations, which then become damped over time. Acoustically, this means that each buckling pulse will cause a longer waveform, which then decays approximately exponentially. Two properties can then be used to characterize such a waveform: the frequency of its wave cycles, and their "quality factor" $Q$. The latter is a scalar measure that quantifies the decrease over time of successive peak amplitudes.

To complicate matters, however, each buckling event modifies the actually resonating mass, and thereby also the frequency and $Q$ value of tymbal plate resonation.

In the algorithm, separate instances of a resonant bandpass filter were used to model the different resonations following each type of buckling pulse. Here, the four center frequencies were explicitly computed from the four mean resonant wave frequencies presented for C. australasiae in [7]; while being scaled to 3× `freq_hz`, so that two decaying wave cycles would fit in between tightly coherent buckling pulses, regardless of the currently required musical pitch. The four corresponding *Q* values for C. australasiae (measured using acoustically unloaded tymbals) were then also used explicitly, to compute the different bandwidths of each filter. Here, care was taken to first halve the *Q* values, so as to make them representative of the acoustically loaded tymbals of an intact cicada.

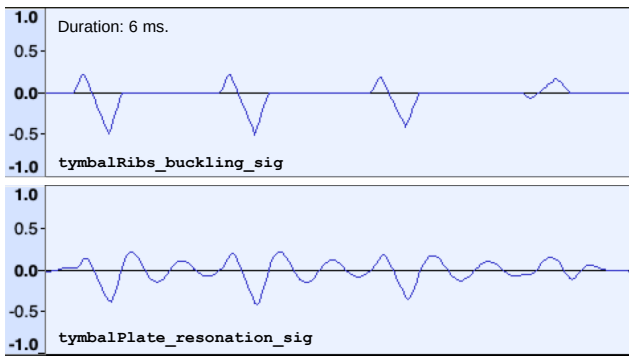

Figure 2. *Top:* example of simulating one complete sequence of tightly synchronized tymbal rib buckling events (at pitch E5). *Bottom:* the corresponding simulated response of the tymbal plate.

The overall `tymbalPlate_resonation_sig` signal was then computed as the weighted sum of the signal containing the simulated buckling pulses and those containing the tymbal plate after-resonations. Here, the weights were set so that the resulting waveform was similar to the acoustically measured results from literature. This included the property that the decayed peak amplitude of the resonant wave cycle occurring directly before the next in-buckling pulse should still be above 25% relative to the maximum – acoustically representing a mechanical vibration amplitude still powerful enough to help start the next tymbal rib's in-buckling (thereby supporting coherence).

The results of this are illustrated in Figure 2, which shows an example tymbal plate resonation waveform caused by a complete sequence of tightly synchronized tymbal rib buckling events, as computed by the algorithm. An example `tymbalPlate_resonation_sig` signal, internally computed for a full cicada call sequence at pitch E5, can be heard in Audio Illustration 3 of the Appendices.

Finally, one speculative addition was made to the computation of tymbal plate resonation. This was based on the impression that perceiving increased loudness during the calls of H. maculaticollis seemed to possibly involve not only an increase in overall amplitude, but also a relative increase of the intensities at higher frequencies. This seemed evident, for example, when listening to the field recording at [5], and then taking two 4-second snippets from its first and second half, amplifying these to the same peak amplitude, and comparing the spectrograms, especially in the 5.0-8.0 kHz range.

This motivated the idea to extend the algorithm with the simulation of some mechanism that would cause its output, too, to include a relative increase at higher frequencies when increasing perceived loudness levels. This seemed desirable not only from a physical modeling perspective, but also from a musical one: The suspected phenomenon seemed reminiscent of how increasing the attack force while playing notes on acoustic musical instruments can have an expressive impact by not just increasing the amplitude, but also the brightness of the tones heard.

The mechanism then concretely modeled was itself highly speculative: Its starting point was that, beyond some level of increasing applied tymbal muscle force, the maximum displacement of the vibrating tymbal plate may become limited to some fixed amount, due to its physical attachment to the rest of the cicada's body. Acoustically, this then *might* result in the wave peaks of tymbal plate resonation becoming progressively flattened as vibration amplitude increases, thereby yielding more intense overtones at higher frequencies. In the algorithm's `tymbalPlate_clippingResonation_sig` signal, the effects of the computations implementing this tentative idea can be observed. However, these effects can also be avoided, easily and completely, by not using the upper half of the `loudness_nrm` input range. Alternatively, the extra computations can be removed from the source code by using an also-included single-line alternative.

### 2.5 The Abdominal Air Sac, Tympana & Opercula: Helmholtz Resonation

As is described in the literature on C. australasiae, movement of the tymbal plate causes air pressure changes going into the abdominal air sac. Together with the left and right tympana – which, variably covered by the opercula, act as the effective sound sources radiating into the external surroundings – this air sac forms an abdominal resonator that appears to produce bottle-shaped or Helmholtz resonation, having the effect of a narrow band-pass filter. Here, the dominant resonant frequency $f_0$ is closely tuned to that of the tymbal, and can be modeled by the equations used in [8] that are listed in (1).

$$\begin{aligned} f_0 &= \frac{c}{2\pi} \cdot \sqrt{\frac{A}{L \cdot V}} \\ A &= 2 \cdot a \\ L &= 16 \cdot r \, / \, 3\pi \\ r &= \sqrt{a \, / \, \pi} \end{aligned} \quad (1)$$

Here, *c* is the speed of sound; *A* is the area of the neck, being twice the area *a* of a single tympanum; *L* is the *effective* length of the neck, determined by *a* via the single-tympanum equivalent hole radius *r*; and *V* is the volume of the cavity. From these dependencies, it then follows that the main variables remaining to be modeled for the case of H. maculaticollis are *V* and *A*, i.e. abdominal volume and effective tympanal area.

**Waveform & spectrogram: field recording** of a complete H. maculaticollis call sequence.   *Duration everywhere: 6.6 seconds.*

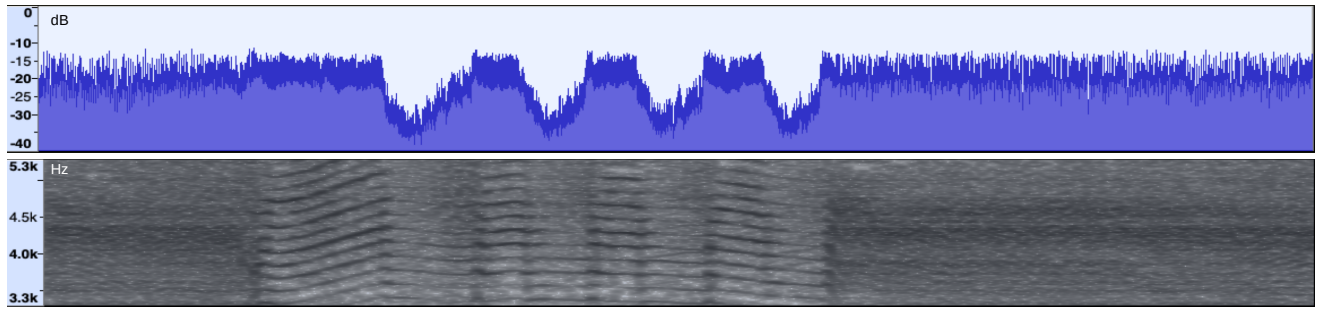

**Waveform & spectrogram: example of corresponding algorithm output** computed for pitch E5.

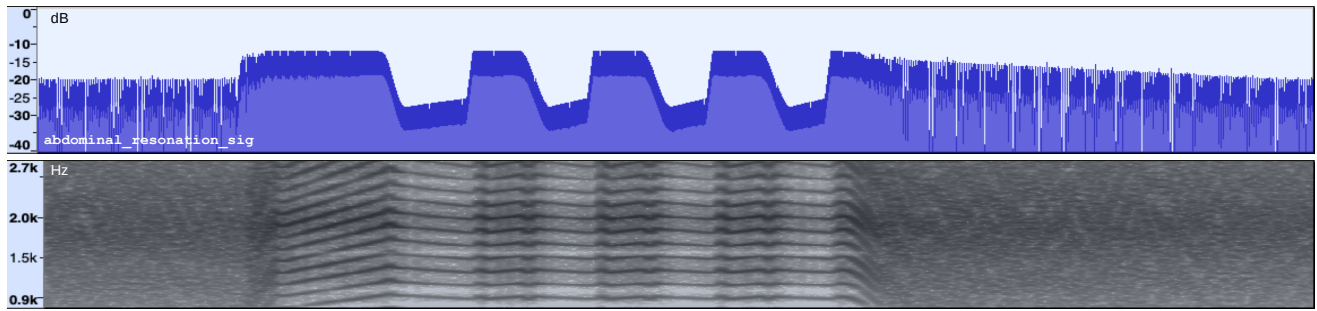

**Waveforms of selected, corresponding internal signals:**

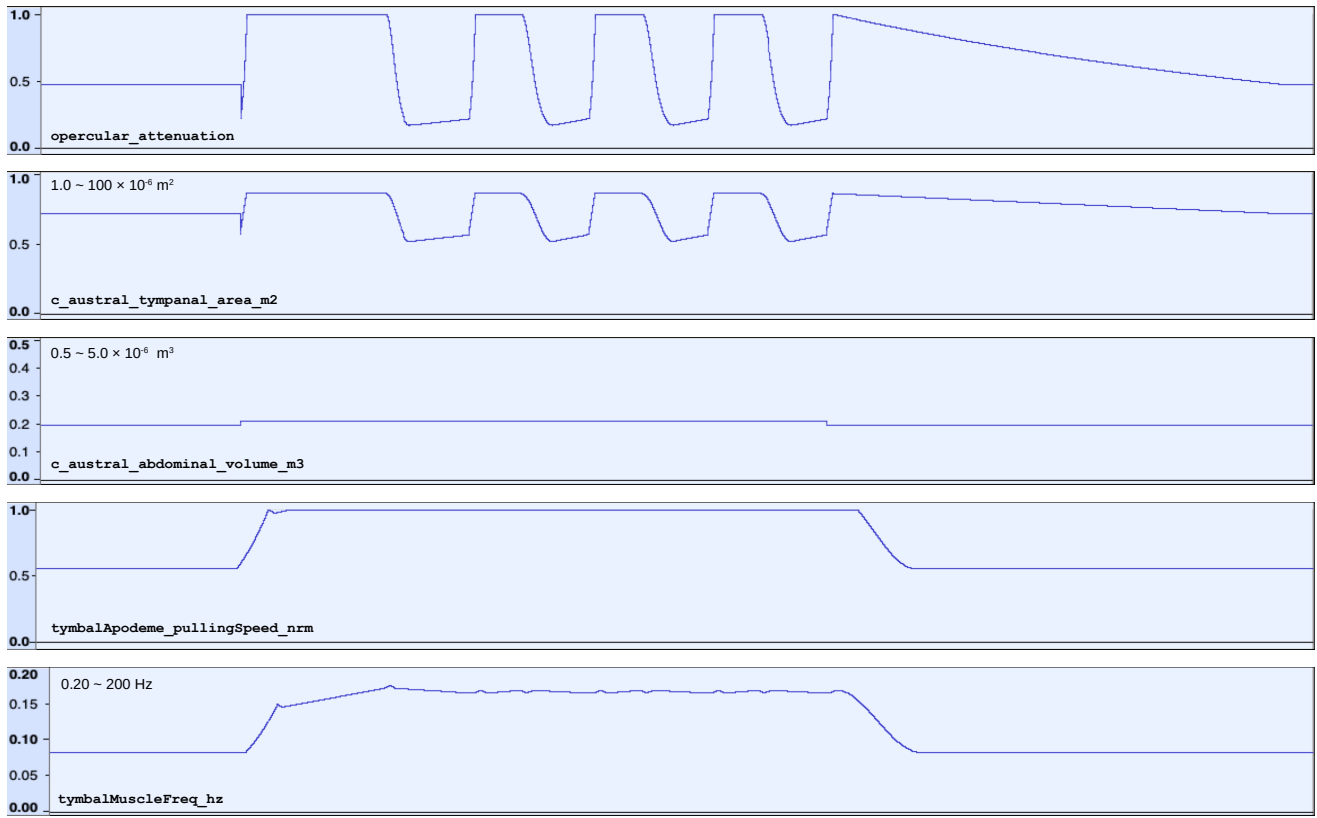

**Waveform of corresponding real-time input signals:**

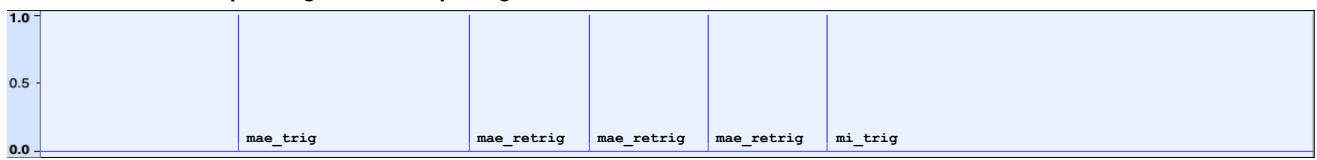

Figure 3.  Algorithmically simulating the calls of H. maculaticollis. Each signal is further explained in the main text.

In the algorithm, abdominal resonation was modeled using a 6-pole (i.e. -36 dB/octave) resonant bandpass filter. Its bandwidth was tuned so as to make algorithm output harmonically similar to H. maculaticollis field recordings, using spectrum plot measurements of the progressive drop in amplitude of the harmonic peaks below insect song frequency (details are in the source code). Fortunately, musically, there was no actual trade-off here when testing at pitch E5, as the resulting bandwidth value seemed to fall within a middle range only outside of which the unintended perception of pitches E6/B6 and E4 seemed to start to occur.

Then, for computing the abdominal volume over time, the starting point were measurements of unextended and extended abdominal cavity volume for C. australasiae, derived from [8]. To this was then added the general idea that H. maculaticollis' abdominal cavity might be less extended during its final call phase. Once modeled, the effects of this seemed hard to discern, however, both via auditory perception and in spectrograms, and the interested reader is therefore referred to the source code for further motivation and details. The resulting computation of `c_austral_abdominal_volume_m3` is illustrated by the waveform placed fourth-lowest in Figure 3.

In comparison, modeling H. maculaticollis' effective tympanal area over time was considerably more involved, once similarly started from the overall tympanal area reported for C. australasiae in [8]. In general, modeling was based on the observation that both cicada species raise their abdomen in some way while singing, thereby lifting the tympana away from the opercula which cover them;[3] and also, on the observation in spectrograms of the H. maculaticollis field recordings that during the initial and middle call phases, this modulation of effective tympanal area seemed to be reliably synchronized with specific patterns in tymbal muscle contraction frequency.

As for tymbal muscle frequency in Section 2.1, a system of envelopes was implemented, one for each insect call phase, with the specific target levels, durations and curvatures defined by a total of 26 scalar constants explicitly based on measurements of amplitude and spectral features present in the field recording obtained from [4]. During this, at three points in the algorithm, musical requirements were consciously prioritized over realistic physical modeling. In each case, envelope segment properties were adjusted so as to perceptually make each simulated insect call phase better correspond to the perceived attack of a single tonal event. This affected 6 constants.[4] The details of all of this are documented in the source code.

A fourth point where musical requirements were prioritized over realistic physical modeling was when recomputing the abdominal resonation frequency resulting from the above relative to the currently intended musical pitch (as specified via `freq_hz`). This, however, will be discussed in the Discussion section.

The resulting computation of `c_austral_tympanal_area_m2` is illustrated by the waveform placed fifth-lowest in Figure 3. Also, an example of the overall resulting `unattenuated_abdominal_resonation_sig` signal can be heard in Audio Illustration 4 of the Appendices, as it was internally computed for a full cicada call sequence at pitch E5.

## 2.6 The Opercula: Amplitude Attenuation

In the literature on C. australasiae, a reduction in sound pressure level is reported when fully closing the opercula. Accordingly, in the algorithm, too, amplitude attenuation of the abdominal resonation signal was implemented, based on the current effective tympanal area. This was done down to a dB SPL level set so that algorithm output matched the drop in strongest harmonic peak amplitude observed during middle call phases in the H. maculaticollis field recording at [4]. An example of the resulting `opercular_attenuation` signal is illustrated by the waveform placed sixth-lowest in Figure 3, as it was internally computed for a full cicada call sequence at pitch E5. The corresponding `abdominal_resonation_sig` signal can be visually compared to the field recording using the spectrograms and waveforms aligned at the top of Figure 3, while it can be heard in Audio Illustration 5 of the Appendices.

## 3. DISCUSSION

From the literature on C. australasiae, it is clear that its abdominal resonation – when abdominal volume is fully extended, and tympanal area fully uncovered – is strongest around the same frequency as tymbal plate resonation. Accordingly, in the algorithm, the abdominal resonation frequency initially was scaled so as to match tymbal plate resonation frequency. This meant that it, too, followed the currently intended musical pitch.

However, when testing this at pitch E4 instead of E5, output from abdominal resonation perceptually seemed to change not only in pitch, but also in timbre, and this to such an extent as to suggest origination from another, different sound source. This seemed problematic, in hindering algorithm output from being perceived as coming from a single source, e.g. during melodic use with other musical voices. An example of this pitch-following abdominal resonation can be heard in Audio Illustration 6 of the Appendices, which contains the `abdominal_resonation_sig` signal for a full call sequence at pitch E4.

Scaling abdominal resonation frequency relative to a *fixed* pitch E5 then seemed to perform better in this regard. In Audio Illustration 7, the corresponding example of fixed-pitch abdominal resonation can be heard. Both options are preserved in the source code.

## 4. CONCLUSION

In Sections 2 and 3, we have discussed the design and implementation of a new algorithm that can simulate the calls of H. maculaticollis, subject to real-time control. Now, revisiting this, as a case study of the possible trade-offs when modeling bioacoustics for tonal use, we can identify 11 instances, extending from the input parameters to the pre-final abdominal resonation stage, where a musical requirement was consciously prioritized over faithful physical modeling.

---

[3] For an H. maculaticollis video example, see:
https://youtu.be/XFOaGPNT55U?t=157.

[4] `c_MAETRIG_TYMPAREA_START_NRM`,
`c_MAETRIG_TYMPAREA_RISE_DUR_S`,
`c_MAERETRIG_TYMPAREA_START_NRM`,
`c_MAERETRIG_TYMPAREA_RISE_DUR_S`,
`c_MITRIG_TYMPAREA_MAX_NRM`,
`c_MITRIG_TYMPAREA_END_NRM`.

This was done for the soprano `freq_hz` input range; the pitch-inducing number of buckling events per muscle contraction; the whole-octave drop in tymbal muscle frequency; the harmonics kept in tune across repeated middle call phases; the pitch-adjusted tymbal plate resonation frequency; the optionally increasing timbral brightness during increasing loudness; the pitch-respecting abdominal resonation bandwidth; the three tympanal area envelope details for better matching insect call phases with single-attack tonal events; and the fixed-pitch abdominal resonation frequency.

This seems to illustrate a number of points that may be relevant when modeling bioacoustics for tonal use. First, that it may be necessary to prioritize musical requirements over faithful physical modeling at many stages of design and implementation. Second, that the resulting adjustments may revolve around having physical modeling perceptually yield sonic events that are well-pitched, single-attack, single-source, and timbrally expressive. Third, that an inherent contradiction may be encountered (here, while modeling abdominal resonation, see Section 3): The pitch-adjusted simulation of resonating bodies may produce sonic results that fail musically precisely when they succeed physically, i.e. when inducing the perception of different sound sources at different pitches.

Finally, the algorithm itself is provided in the Appendices of this paper, together with the audio illustrations that have already been used to demonstrate the computation of some of its key internal signals. In addition to this, Audio Illustration 8 demonstrates the tonal use of overall algorithm output together with another instrument, containing brief examples of counterpoint and chordal accompaniment by a digital piano. Throughout the algorithm's source code, emphasis has been put on readability and ease of reuse: The overall structure is flat and unambiguous, and uses human-readable constants and variables together with detailed in-line comments that explicitly motivate each of the choices made. The intention here was to make it easier for (in principle) anybody to use, modify or extend the algorithm. This could include changes to more faithfully simulate the calls of cicadas in general (e.g., by explicitly modeling a *pair* of tymbals); changes to better match H. maculaticollis, C. australasiae, or any other species of cicada; and changes to the trade-offs between realistic physical modeling and musical use.

### Acknowledgments

I would like to take the opportunity to thank my hosts in Gyeongju for their sincere hospitality. Visiting this region formed a great part of my first (and hopefully not last) stay in the Republic of Korea, during which I first heard the calls of H. maculaticollis in real life.

## 5. APPENDICES

The appendices are included inside the PDF of this paper. Each embedded file can be opened or exported using its (shortened) name in the text. For the Audio Illustrations, this is via: "Audio Illustration 1", "2", "3", "4", "5", "6", "7", and "8". For the source code of the algorithm, this is via: "`SoundGeneratingProcesses.sc`". The latter needs to be placed among the Extensions of the SuperCollider installation that is then used to run the code (details provided). Finally, example source code that instantiates and parametrizes the algorithm is provided via: "`usage example.scd`".